\documentclass[a4paper,11pt]{article}
\pdfoutput=1 % if your are submitting a pdflatex (i.e. if you have
\usepackage{jinstpub} % for details on the use of the package, please
\usepackage{lineno}
\title{\boldmath BLEMAB European project: muon imaging technique applied to blast furnaces}

\author[a,1]{L. Bonechi,\note{Corresponding author.}}
\author[b,c]{F. Ambrosino,}
\author[d]{P. Andreetto,}
\author[e,f]{G. Bonomi,}
\author[a,g]{D. Borselli,}
\author[a]{S. Bottai,}
\author[h]{T. Buhles,}
\author[i]{I. Calliari,}
\author[d]{P. Checchia,}
\author[j]{U. Chiarotti,}
\author[a]{C. Cialdai,}
\author[a]{R. Ciaranfi,}
\author[b,c]{L. Cimmino,}
\author[k,a]{V. Ciulli,}
\author[k,a]{R. D’Alessandro,}
\author[b,c]{M. D’Errico,}
\author[l]{R. Ferretti,}
\author[h]{F. Finke,}
\author[h]{A. Franzen,}
\author[m]{B. Glaser,}
\author[k,a]{S. Gonzi,}
\author[m]{Y. Liu,}
\author[n,d]{A. Lorenzon,}
\author[c]{V. Masone,}
\author[o]{O. Nechyporuk,}
\author[i]{L. Pezzato,}
\author[m]{B.V. Rangavittal,}
\author[p]{D. Ressegotti,}
\author[b,c]{G. Saracino,}
\author[h]{J. Sauerwald,}
\author[a]{O. Starodubtsev,}
\author[a]{and L. Viliani}

\affiliation[a]{INFN, Florence, Italy}
\affiliation[b]{Physics Department "Ettore Pancini", University of Naples Federico II, Naples, Italy}
\affiliation[c]{INFN, Naples, Italy}
\affiliation[d]{INFN, Padua, Italy}
\affiliation[e]{Department of Mechanical and Industrial Engineering, University of Brescia, Brescia, Italy}
\affiliation[f]{INFN, Pavia, Italy}
\affiliation[g]{Department of Physics and Geology, University of Perugia, Perugia, Italy}
\affiliation[h]{Ironmaking Department, ArcelorMittal Bremen GmbH, Bremen, Germany}
\affiliation[i]{Department of Industrial Engineering, University of Padua, Padua,Italy}
\affiliation[j]{Rina Consulting – Centro Sviluppo Materiali SpA, Rome, Italy}
\affiliation[k]{Department of Physics and Astronomy, University of Florence, Florence, Italy}
\affiliation[l]{Rina Consulting – Centro Sviluppo Materiali SpA, Terni, Italy}
\affiliation[m]{KTH Royal Institute of Technology, Stockholm, Sweden}
\affiliation[n]{Department of Physics and Astronomy, University of Padua, Padua, Italy}
\affiliation[o]{ArcelorMittal Maizieres Research SA, Maizieres -Les-Metz, France}
\affiliation[p]{Rina Consulting – Centro Sviluppo Materiali SpA, Dalmine (BG), Italy}

\emailAdd{lorenzo.bonechi@fi.infn.it}

\abstract{The BLEMAB European project (BLast furnace stack density Estimation through on-line Muon ABsorption measurements), evolution of the previous Mu-Blast European project, is designed to investigate in detail the capability of muon radiography techniques applied to the imaging of a blast furnace's inner zone. In particular, the geometry and size of the so called "cohesive zone", i.e. the spatial zone where the slowly downward moving material begins to soften and melt, that plays an important role in the performance of the blast furnace itself. Thanks to the high penetration power of the natural cosmic ray muon radiation, muon transmission radiography represents an appropriate non-invasive methodology for  imaging large high-density structures such as blast furnaces, whose linear size can be up to a few tens of meters. A state-of-the-art muon tracking system, whose design profits from the long experience of our collaboration in this field, is currently under development and will be installed in 2022 at a blast furnace on the ArcelorMittal site in Bremen (Germany) for many months. Collected data will be exploited to monitor temporal variations of the average density distribution inside the furnace. Muon radiography results will also be compared with measurements obtained through an enhanced multipoint probe and standard blast furnace models.}


 \collaboration{\includegraphics[height=17mm]{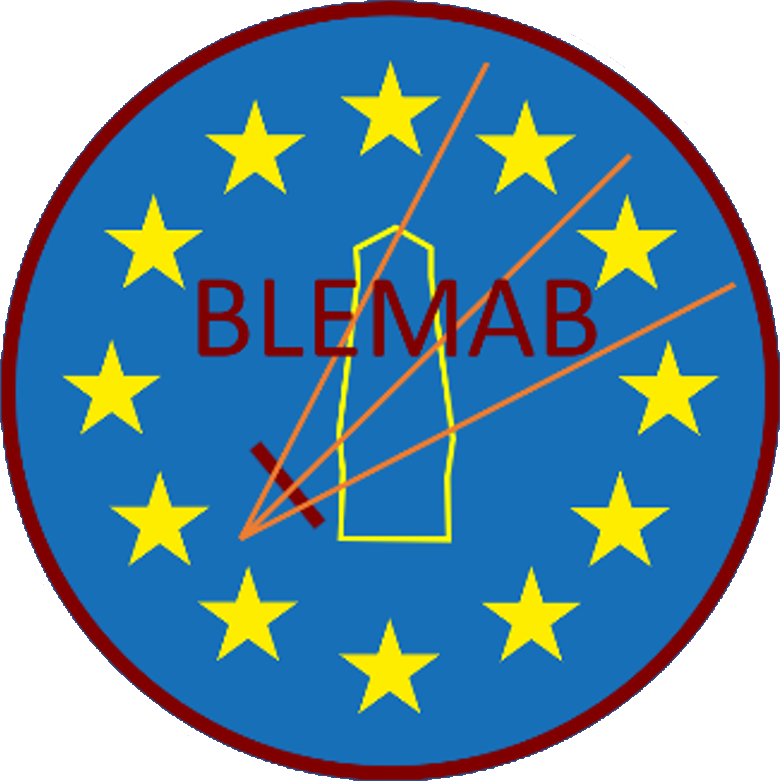}\\[6pt]
   BLEMAB collaboration}
\proceeding{22$^{\text{nd}}$ International Workshop on Radiation Imaging Detectors\\
  27 June 2021 - 01 July 2021\\
    Ghent, Belgium (online)}

\begin{document}
\maketitle
\flushbottom

%********************************************************************
%********************************************************************
%**************             INTRODUCTION              ***************
%********************************************************************
%********************************************************************
\section{Introduction}
\label{sec:intro}
Based on the usage of the natural muon flux continuously produced by high energy cosmic rays in the Earth's atmosphere (see for example \cite{PDG}), muon transmission radiography is a non-invasive survey technique that can be profitably used in many fields, from Archaeology to civil engineering, from mining to volcanology. This technique provides precise information on the two-dimensional average density distribution of very thick material layers up to hundreds of meters thick. For this reason it represents a powerful tool complementary or alternative to other more common survey methods, in particular in those cases where the material thicknesses are too large or the volume to be measured not accessible.

The next two paragraphs provide an introduction to the muon transmission radiography technique and to the BLEMAB project, the purpose of which is mainly the validation of the muon imaging methodology in an important industrial context.

%--------------------------------------------------------------------
%--------------      Muon Radiography         -----------------------
%--------------------------------------------------------------------
\subsection{Muon radiography}
\label{sec:intro:muon_radiography}

Like electrons, muons are charged elementary particles belonging to the lepton family, a type of particle that can interact through all types of forces except the hadronic force. Muons are continuously produced in the upper layer of the Earth's atmosphere due to the collisions of very high-energy cosmic rays with atmospheric gas. Due to their high energies and their large mass, about 200 times that of the electron, atmospheric muons can reach the earth's surface and penetrate up to hundreds of meters of rock, depending on their energy.

The basic idea of muon radiography is to exploit the natural atmospheric muon flux to perform radiographic studies of large massive structures, thus reproducing in a very similar way what is usually done with x-rays in the usual radiographs.
The most direct result of a muon radiograph is the measurement of the directional transmission of the muons, that is, the fraction of incoming muons that is able to cross the thickness of the material encountered along one direction. Transmission can be experimentally estimated by comparing the muon flux measured downstream of the target under examination with the flux measured upstream in the same direction. The interpretation of the measured transmission requires a comparison with simulations that takes into account the geometry of the volume under examination, the characteristics of the muon flux and the other information available on the material structures placed in the acceptance of the experimental apparatus used for the measurement. Many simulations, algebraic formulas and experimental measurements can be found in the literature describing muon flux at ground level, but often these are not sufficient to fully describe the energy and angular ranges required for muon radiography applications. Atmospheric muon flux has important dependencies both on energy and direction. The INFN-UNIFI team in Florence measured muon energy and angular spectra at ground level in the momentum range from 0.1 to 130 GeV/$c$ and in the zenith angle range from $0^{\circ}$ up to 80$^{\circ}$ using the ADAMO detector \cite{BonechiL_ADAMO_2005}, a magnetic spectrometer made with a permanent magnet and a microstrip silicon tracker, with a Maximum Detectable Rigidity of about 250\,GeV/$c$. Based on this measurement, a preliminary version of a muon software generator was implemented and successfully used to compare different muon radiographs with dedicated simulations \cite{SaracinoG_GALLERIABORBONICA2017,BonechiL_ICNFP2017,BaccaniG_ICNFP2018,BonechiL_IPRD2019,BaccaniG_ARGINI2021}. ADAMO results have been also  used recently for the implementation of a realistic atmospheric muon generator \cite{PaganoD_ECOMUG_2021} integrated with the GEANT4 software package \cite{GEANT4}.

The comparison of the measured  two-dimensional angular distribution of muon transmission  with realistic software simulations allows reconstructing the two-dimensional average angular density distribution of the target, as resulting from the measuring position. Combining multiple measurements performed from different viewpoints, information on the three-dimensional density distribution can be derived. 

Detailed reviews about muon imaging and the relative instrumentation can be found in \cite{BonomiG_Review2020,BonechiL_Review2020}.

%--------------------------------------------------------------------
%--------------      The BLEMAB project      ------------------------
%--------------------------------------------------------------------
\subsection{The BLEMAB project}
\label{sec:intro:project}

The main aim of the BLEMAB project is to test the capability of muon transmission radiography technique in internal volume imaging of blast furnaces and to establish a non-invasive investigation methodology for online monitoring of internal density variations of such structures.

An important test of the performance of this imaging technique concerns the ability to measure the shape, position and thickness of the so-called "cohesive zone", the place in the blast furnace where the iron-containing raw material begins to melt.
The knowledge of the processes that take place inside the blast furnaces and the maximization of the extension of the cohesive zone are crucial points for improving the productivity of these plants and therefore of great interest for steel companies.
Until now, this type of study has been based only on direct measurements, such as the tracking of radioisotopes injected from the top of the blast furnace or the use of vertical probes with conductive cables connected to conventional instrumentation, such as vertical multi-point probes (MPVP), which provide essential information about the state of the internal volume \cite{L03,L04}, or other very expensive and invasive \cite{L05} techniques.
Unlike these techniques, muon imaging is a safe non-invasive radiographic methodology based on a completely natural and freely usable radiation, which therefore does not even cause any radiological issue.

Muon imaging has been proposed in the past for blast furnace imaging, as described in \cite{MUBLAST,L09}. These works demonstrated the feasibility of measuring the density distribution of the lower part of a blast furnace using the muon radiography technique. However, a direct measurement of the height of the casting surface had not yet been hypothesized. A simple muon detection system has already been employed by AM to verify the possibility of detecting density variations inside a BF's inner volume by studying the corresponding variation in the flux of atmospheric muons crossing it. The study of the average density distribution inside the BF would add important hints about the position and extension of the cohesive zone as a function of the BF's running parameters. According to the BF's experts the measurement of average density variations in some regions of the BF's inner volume in a timescale of some days after changing the running parameters would be extremely useful for improving the BF's steel production process. 

This is the main objective of the BLEMAB proposal, for which a muon tracking system of large acceptance and appropriate angular resolution is being developed. This equipment will be installed at the blast furnaces hosted in the ArcelorMittal (AM) site in Bremen (Germany) for many months, in order to allow long-term data collection and the development of a monitoring methodology based on the online analysis of muon trajectories and on the comparison with realistic simulations. Muographic measurements will also be compared with measurements obtained using an enhanced multipoint vertical probe (E-MPVP) and standard blast furnace models for validation.

Section \ref{sec:simulations} introduces the software simulation framework used for a preliminary simulation and a new framework based on the GEANT4 package, already available but still under development for further improvements, that will be used to realize the full simulations using the most realistic models of the AM's blast furnaces.

Some details of the prototype detector under development for the BLEMAB project are presented in section \ref{sec:apparatus}, together with preliminary results obtained from a test of a small prototype tracking plane implementing the BLEMAB concepts.

%********************************************************************
%********************************************************************
%**************       SOFTWARE SIMULATIONS            ***************
%********************************************************************
%********************************************************************
\section{Software simulations}
\label{sec:simulations}
Software simulations are required essentially for two main purposes:
\begin{itemize}
    \item first, to provide preliminary estimates of the expected muon rate, in such a way to define the minimum size of the apparatus to have a statistically relevant measurement in time intervals of the order of 24\,h;
    \item later, to perform as precise as possible comparisons with real measurements, allowing to convert the two-dimensional muon transmission maps in average density maps.
\end{itemize}
Two different kind of implementations are considered for the BLEMAB project: a simplified custom one, which is used for all preliminary or quick evaluations, and a full GEANT4 based simulation for more realistic evaluations and final comparison with results. Results obtained using the former one are shown in paragraph \ref{sec:simulations:simple}, while a short introduction to the latter one is presented in paragraph \ref{sec:simulations:geant4}.

%--------------------------------------------------------------------
%--------------      Simplified simulation     ------------------------
%--------------------------------------------------------------------
\subsection{Simplified implementation}
\label{sec:simulations:simple}
The main evaluations concerning the feasibility of the experiment have been finalized thanks to a simulation tool already available to the team and successfully used for previous studies related to measurements carried out at the Temperino mine in Campiglia Marittima, near Livorno (Italy) \cite{BonechiL_ICNFP2017,BaccaniG_ICNFP2018,BonechiL_IPRD2019}, at the Bourbon Gallery in Naples \cite{SaracinoG_GALLERIABORBONICA2017}  (Italy) and at river embankments near Florence and Pistoia \cite{,BaccaniG_ARGINI2021}  (Italy). In general all simulations for muographic applications require the knowledge of the target material thickness $l(\theta,\phi)$ and its average density $\rho(\theta,\phi)$ as seen from the hypothetical detector's installation position along all directions in its field of view, defined by the zenith ($\theta$) and azimuth ($\phi$) angles. The product of these two quantities defines the target material opacity along each direction, $X(\theta,\phi)=l(\theta,\phi)\times\rho(\theta,\phi)$, which in turn determines the minimum momentum $p_{min}(\theta,\phi)$ that a muon coming from that direction must have to cross both the target material and the detector tracking planes, to be detected. In this simulation $p_{min}(\theta,\phi)$ is determined by using tabulated muon range values available in the literature. The corresponding muon transmission $t(\theta,\phi)$ is finally calculated as the ratio between the integral of the muon momentum spectrum $\Phi(p,\theta,\phi)$ (for which analytical expressions have been fitted to the ADAMO data) in the momentum range beyond $p_{min}(\theta,\phi)$ (simulating a measurement performed downstream the target volume) and the integral of the same spectrum in the whole momentum range down to the minimum value allowing muons to cross the detector modules (simulating a free-sky reference measurement).

For the BLEMAB application a 45$^{\circ}$ tilt of the detector pointing direction with respect to the vertical direction is assumed for pointing the blast furnace's cohesive zone; in this configuration a total muon rate of approximately 40\,Hz is expected with a single BLEMAB detector.

\begin{figure}[t!]
\begin{center} %takes some additional vertical space
\includegraphics[height=.21\textwidth]{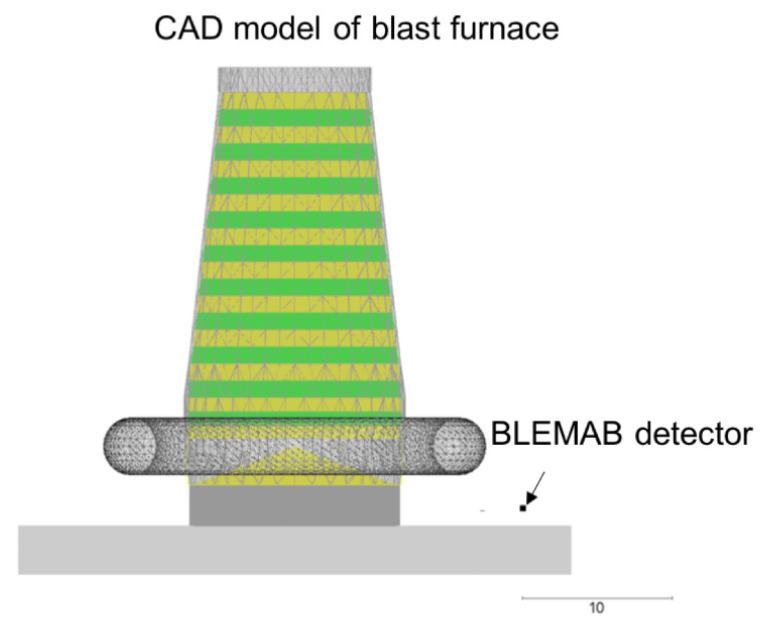}
%\hspace{.0\textwidth}
\includegraphics[height=.21\textwidth]{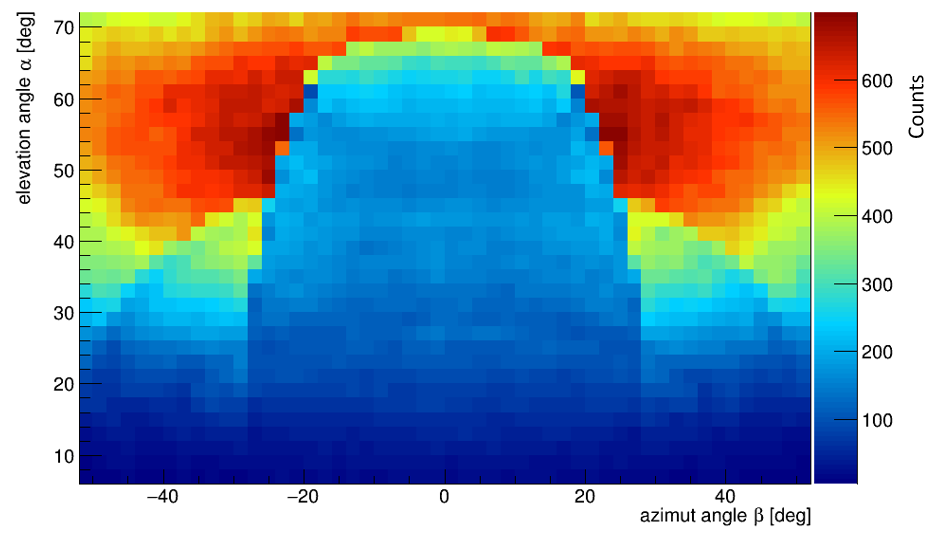}
%\hspace{.0\textwidth}
\includegraphics[height=.21\textwidth]{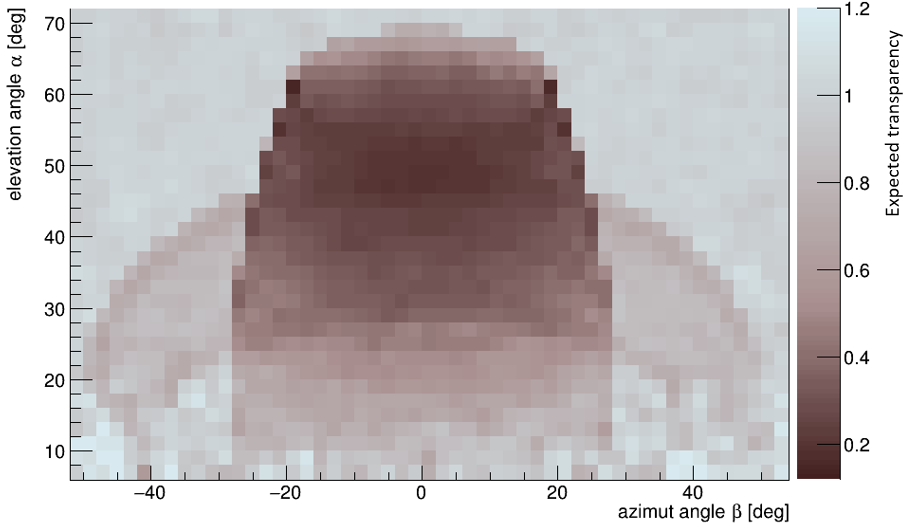}
\end{center}
\vspace{-0.8cm}
%\qquad
%\includegraphics[height=.27\textwidth,origin=c]{figures/TJKL7050.JPG}
% "\includegraphics" from the "graphicx" permits to crop (trim+clip)
% and rotate (angle) and image (and much more)
\caption{\label{fig:preliminary_stand-alone-simulation}  {\bf Left}: sketch of the geometric configuration implemented for the preliminary simulations, consisting of a cubic detector with a 80\,cm$\times$80\,cm incoming surface, tilted 45$^{\circ}$ with respect to the zenith and placed 8\,m far from the wall of a structure reproducing the blast furnace geometry and material. {\bf Center}: two dimensional angular distribution of muon events expected after approximately 10\,h data taking. The shadow of the blast furnace is clearly visible in the central part of the plot,for azimuth angle between -30$^{\circ}$ and +30$^{\circ}$ and elevation below 70$^{\circ}$, where the number of detected muon is clearly suppressed. {\bf Right}: simulated muon transmission measurement expected with a 10\,h data taking.}
\end{figure}

The sketch on the left in figure \ref{fig:preliminary_stand-alone-simulation} shows the geometry considered for this simulation, implementing a simplified model of the blast furnace located in the Bremen site of ArcelorMittal.The two dimensional angular distribution of muon events expected downstream the blast furnace is show at center, considering 10\,h muon events approximately. This distribution is based on a real free sky measurement performed with the MIMA detector and on the simulated muon transmission distribution, reported in the plot on the right side.

In this simulation a 2$^{\circ}$ bin size is chosen both for elevation and zenith angles. The presence of the blast furnace in front of the muon detection system can be identified as an elongated angular region in the central part of the distributions, where the number of detected muons is significantly suppressed. 

In each bin two to three hundred muon tracks are expected after a data taking of about 24\,h in the angular region where the cohesive zone is located, allowing a measurement of muon distribution with statistical fluctuations below 10\,\% with such an angular resolution.

Yet more precise information can be obtained integrating muon events over larger angular regions. Statistical fluctuations at percent level in half day are expected when observing a $10^{\circ}\,\times\,10^{\circ}$ region in the direction of the cohesive zone.

\begin{figure}[t!]
\begin{center} %takes some additional vertical space
\includegraphics[height=.21\textwidth]{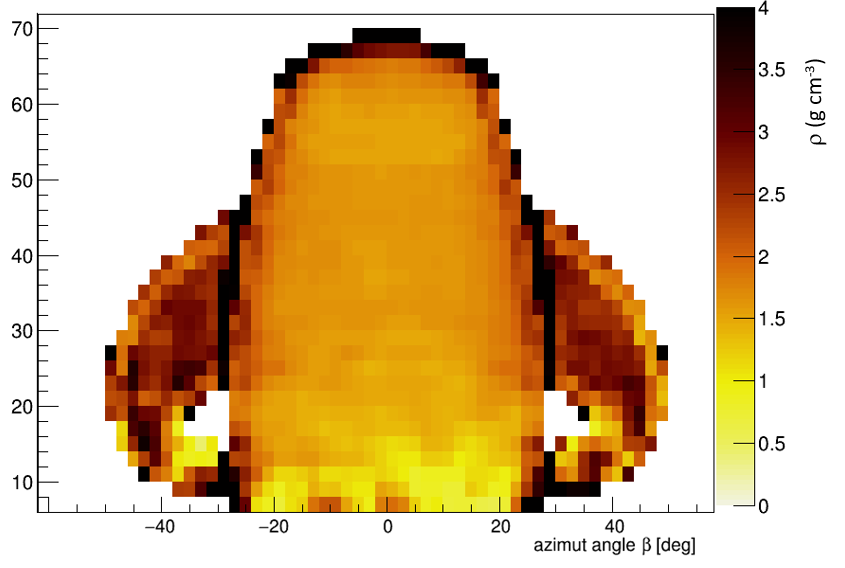}
%\hspace{.0\textwidth}
\includegraphics[height=.21\textwidth]{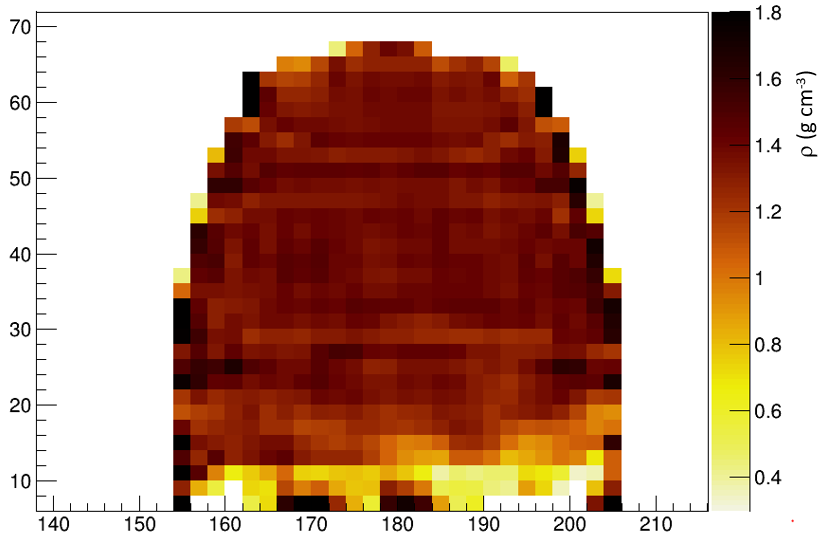}
%\hspace{.0\textwidth}
\includegraphics[height=.21\textwidth]{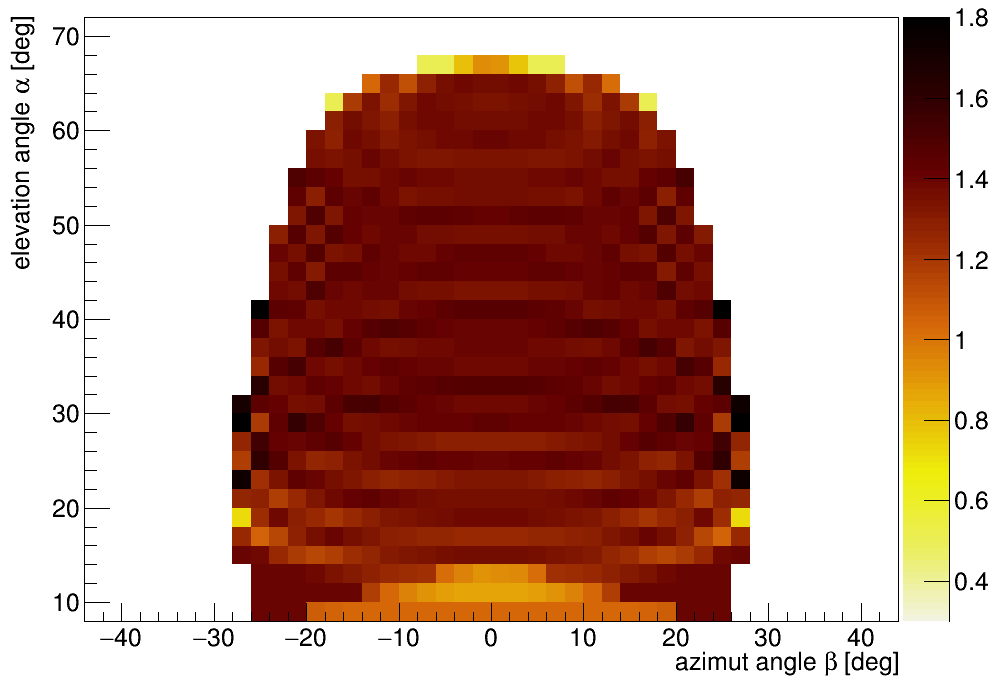}
\end{center}
\vspace{-0.8cm}
\caption{\label{fig:simple-simulation:density} Main results of the BLEMAB stand-alone custom simulation. The three plots report (left) the expected average density angular distribution reconstructed for the full structure, (center) the same plot after subtracting the contribution due to the toroidal structure surrounding the blast furnace and to the outer metal shell and (right) the average density distribution implemented in the simulation.}
\end{figure}

Figure \ref{fig:simple-simulation:density} shows the main results of this simulation. In particular, the expected two dimensional angular distribution of the average density has been evaluated for the whole blast furnace structure and after subtracting the contribution due to the toroidal structure surrounding the furnace and to the outer metal shell. The angular average density distribution implemented in the simulation is shown on the right side plot for comparison. This plots show that some details of the internal material distribution can be argued already after a 10\,h data taking, but is must be observed that in reality the material layers inside a blast furnace are constantly falling down to the melting zone, taking approximately 8\,h to complete a full cycle. Therefore an average over time of the average material density will be measured in this time interval. In a more realistic scheme, the difference in material density in the lower and upper parts of the blast furnace's volume is expected to be larger (0.7\,g\,cm$^{-3}$ versus 1.5\,g\,cm$^{-3}$) and the interface between the two regions should be yet more evident.

%--------------------------------------------------------------------
%--------------      The full simulation     ------------------------
%--------------------------------------------------------------------
\subsection{Implementation of a full simulation}
\label{sec:simulations:geant4}
A more detailed simulation has been developed, which implements an accurate representation of the geometry and material composition of the blast furnace where the measurement will be carried out. The new simulation is based on the GEANT4 software package and on a recently developed atmospheric muon software generator \cite{PaganoD_ECOMUG_2021}. The basic structure of this simulation was derived from the one created previously within the European MU-BLAST project \cite{MUBLAST}.

\begin{figure}[htbp]
\centering
\includegraphics[width=.85\textwidth]{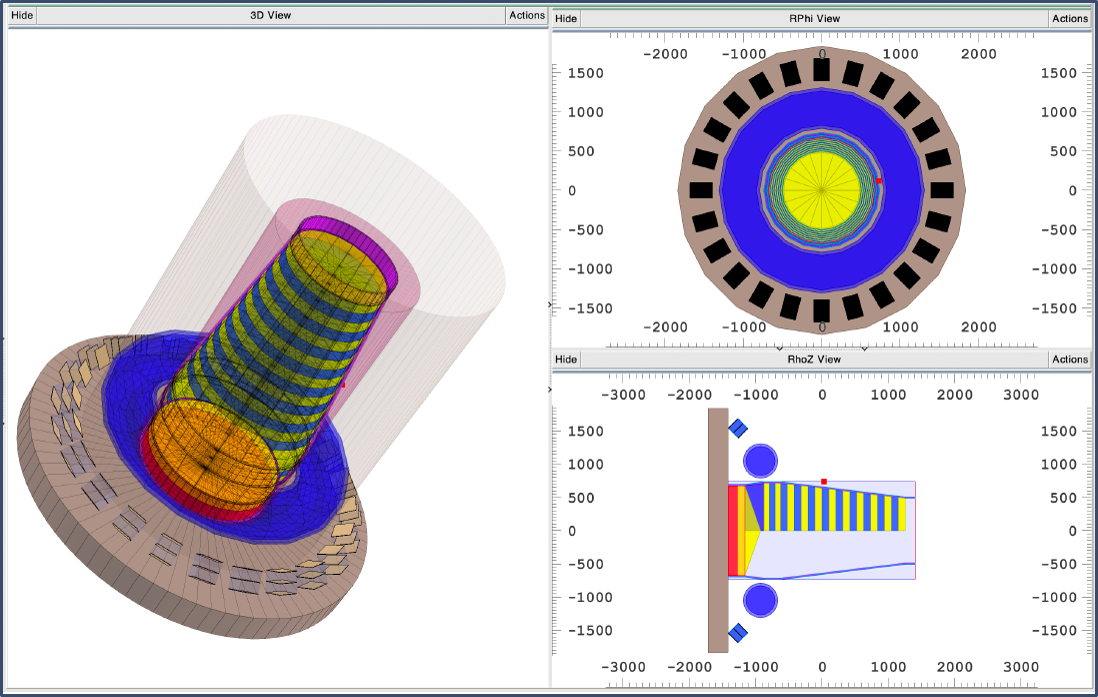}
\caption{\label{fig:full_simulation} Implementation of geometry and materials in the detailed BLEMAB simulation framework under development, based on the GEANT4 software package. The three boxes show three different views of the system. Many identical detectors, each composed of three tracking planes, have been placed around the structure in such a way to maximize the number of simulated events that it is possible to obtain from the simulation, for which a software muon generator based on the ADAMO atmospheric muon data at ground level is also implemented.}
\end{figure}

Figure \ref{fig:full_simulation} shows a sketch of the geometry considered for the full simulation of the BLEMAB experimental configuration.

In this simulation 24 identical detectors, 80\,cm\,$\times$\,80\,cm\,$\times$\,80\,cm are placed all around the blast furnace, 8\,m far from the wall and some meters below the tuyere, the large toroidal pipe surrounding the structure, used to blow hot air inside the blast furnace. The multiple detection systems allow increasing the ratio between the number of events entering the detector's acceptance and the total number of simulated muon events, which can be generated from a cylindrical surface surrounding the blast furnace.

Results from this complete simulation, which takes into account a detailed description of physics interactions of muons inside materials, will be used for more careful evaluations and for comparisons with the experimental measured distributions.
%********************************************************************
%********************************************************************
%**************               APPARATUS               ***************
%********************************************************************
%********************************************************************
\section{Experimental apparatus}
\label{sec:apparatus}
The application of the muon transmission radiography survey method requires the installation of a muon tracking system downstream of the volume of interest. For the BLEMAB project two independent muon trackers will be realized in order to have the possibility of installing them a) both in the same blast furnace, for a stereoscopic view of the structure, or b) in two different blast furnaces at the same time, to allow the comparative study of different structures, or c) one in front of a blast furnace and the other in an open space for a reference measurement.

Each of these trackers is made of three independent 80\,cm$\times$80\,cm tracking modules measuring two coordinates of muon impact points along two orthogonal axes (XY). The tracking modules are based on the same technology used for the MURAVES \cite{DErricoME_IPRD2019} and MIMA \cite{BaccaniG_MIMA2018} projects.

Each XY tracking module is composed of two stacked tracking planes each of which consists of 64 scintillator bars,  80\,cm long, with triangular section. The orientation of bars in the two planes are orthogonal to each other. 

The scintillator material chosen for the BLEMAB detectors is a good quality fast organic polystyrene based plastic, characterized by large bulk attenuation length, large light yield and small signal rise time. Each scintillator bar is read by means of two silicon photomultipliers 4\,$\times$\,4\,mm$^2 $ with optimized quantum efficiency in correspondence with the emission spectrum of the selected scintillator.

Each tracking module is readout by means of four custom DAQ slave boards housing a EASIROC1B 32 channels front-end chip and an ASIC to control the front-end circuitry and the transmission of data to a central custom DAQ master board implementing the trigger logic and data collection from up to 16 slave boards. A Raspberry PI miniaturized computer, connected to the master DAQ board, is used to setup the whole electronic chain, receive the data packet from the master DAQ board and write it to a physical support. A network access to the detector will be available at the AM blast furnace to allow online detector control and a synchronization of data to a remote server.

\begin{figure}[t!]
\centering % \begin{center}/\end{center} takes some additional vertical space
\includegraphics[height=.3\textwidth,]{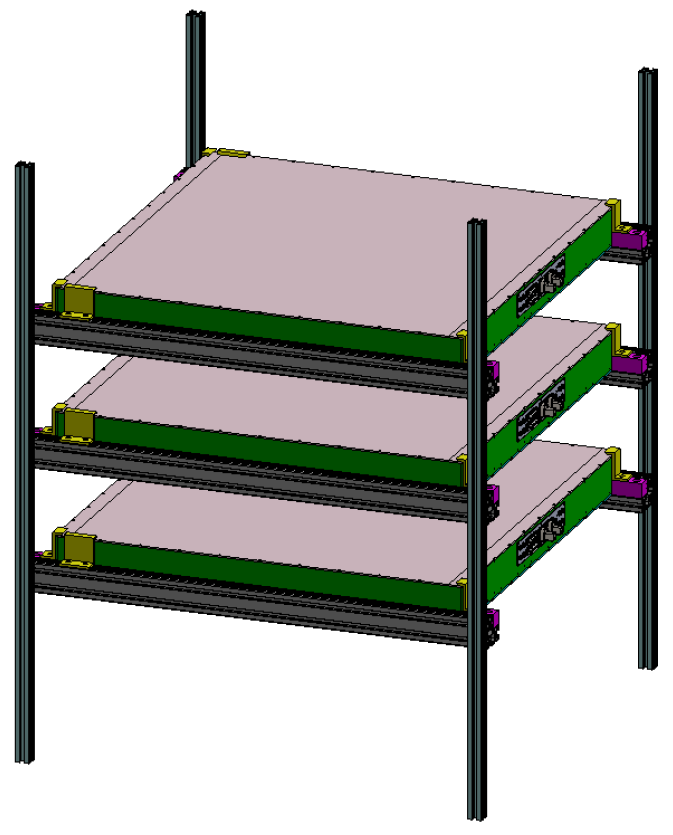}
%\hspace{3mm}
\qquad
\includegraphics[height=.3\textwidth,]{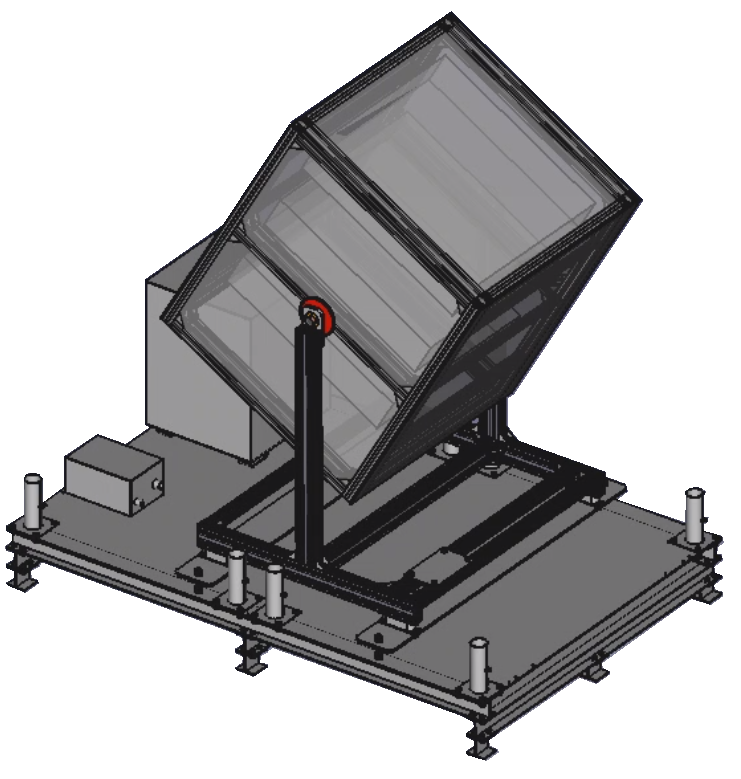}
%\hspace{3mm}
\qquad
\includegraphics[height=.3\textwidth,]{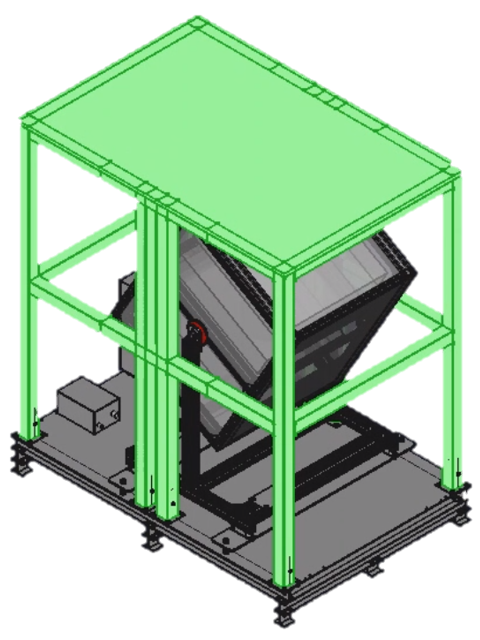}
% "\includegraphics" from the "graphicx" permits to crop (trim+clip)
% and rotate (angle) and image (and much more)
\caption{\label{fig:mechanics} Left: single BLEMAB tracker, composed of three XY tracking modules. Center: the BLEMAB tracker installed on its mounting and placed on the base plate. Right: skeleton of the protection frame with a steel cover plate at its top (in green). The remaining metal plates have been hidden in this drawing.}
\end{figure}

A single muon tracker, an assembly of three XY tracking modules, will be enclosed in an light aluminum box and fixed on a custom mounting which allows to modify the detector pointing direction. The whole measuring system will be housed inside a large metal frame, completely covered by metal panels, to protect the apparatus from possible blows, splashes of liquids and corrosive vapors. Some technical drawings are shown in figure \ref{fig:mechanics}.

Because the blast furnace's environment is usually characterized by quite high temperatures, a cooling system based on a water chiller will be housed inside the same protection frame and the cooling line will be put in contact with the detector's box, in order not to exceed the maximum working temperature of the optical sensors. 

%--------------------------------------------------------------------
%--------------      Test of prototype       ------------------------
%--------------------------------------------------------------------
\subsection{Construction and test of a small prototype tracking plane}
\label{sec:apparatus:prototype}
\begin{figure}[t!]
\centering
\includegraphics[height=.21\textwidth,trim=30 110 0 0,clip]{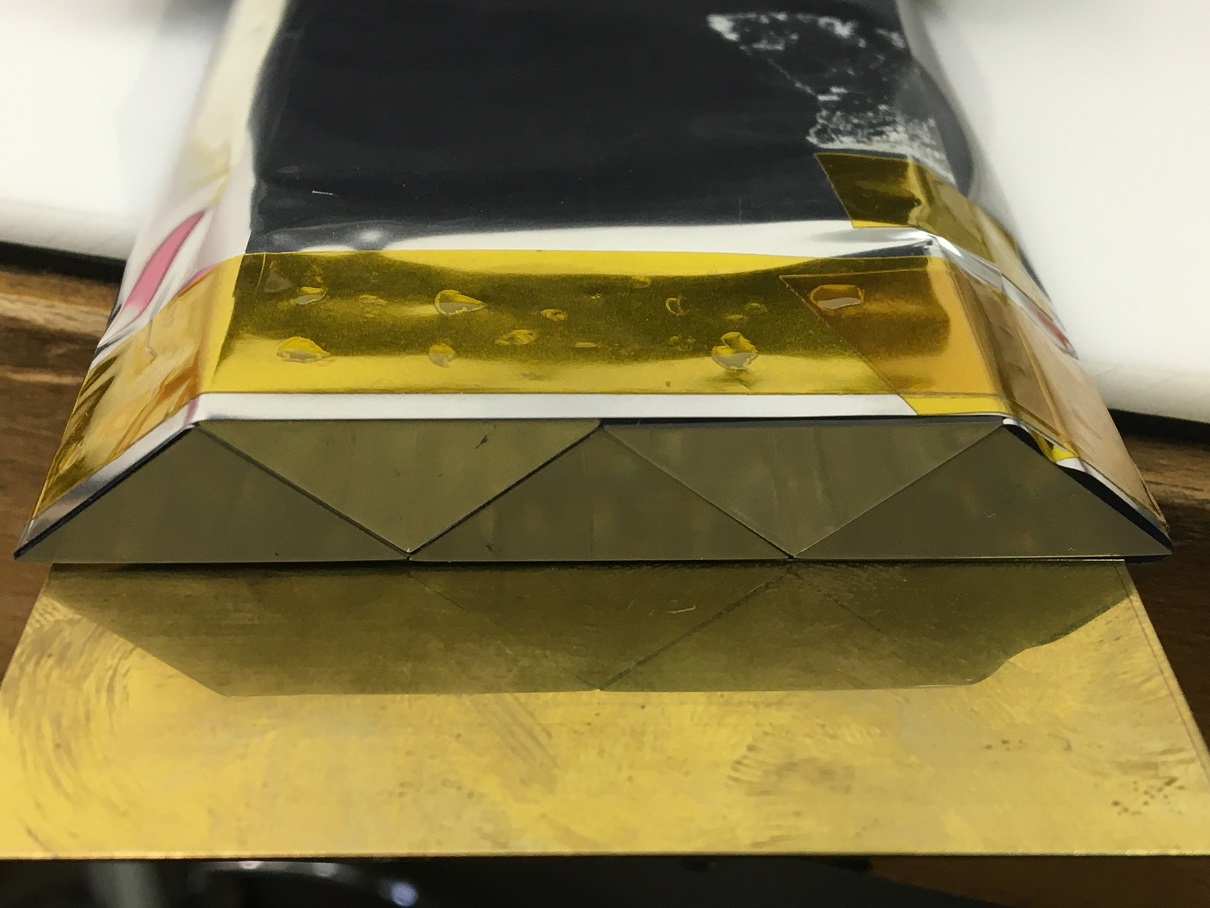}
\includegraphics[height=.21\textwidth,origin=c]{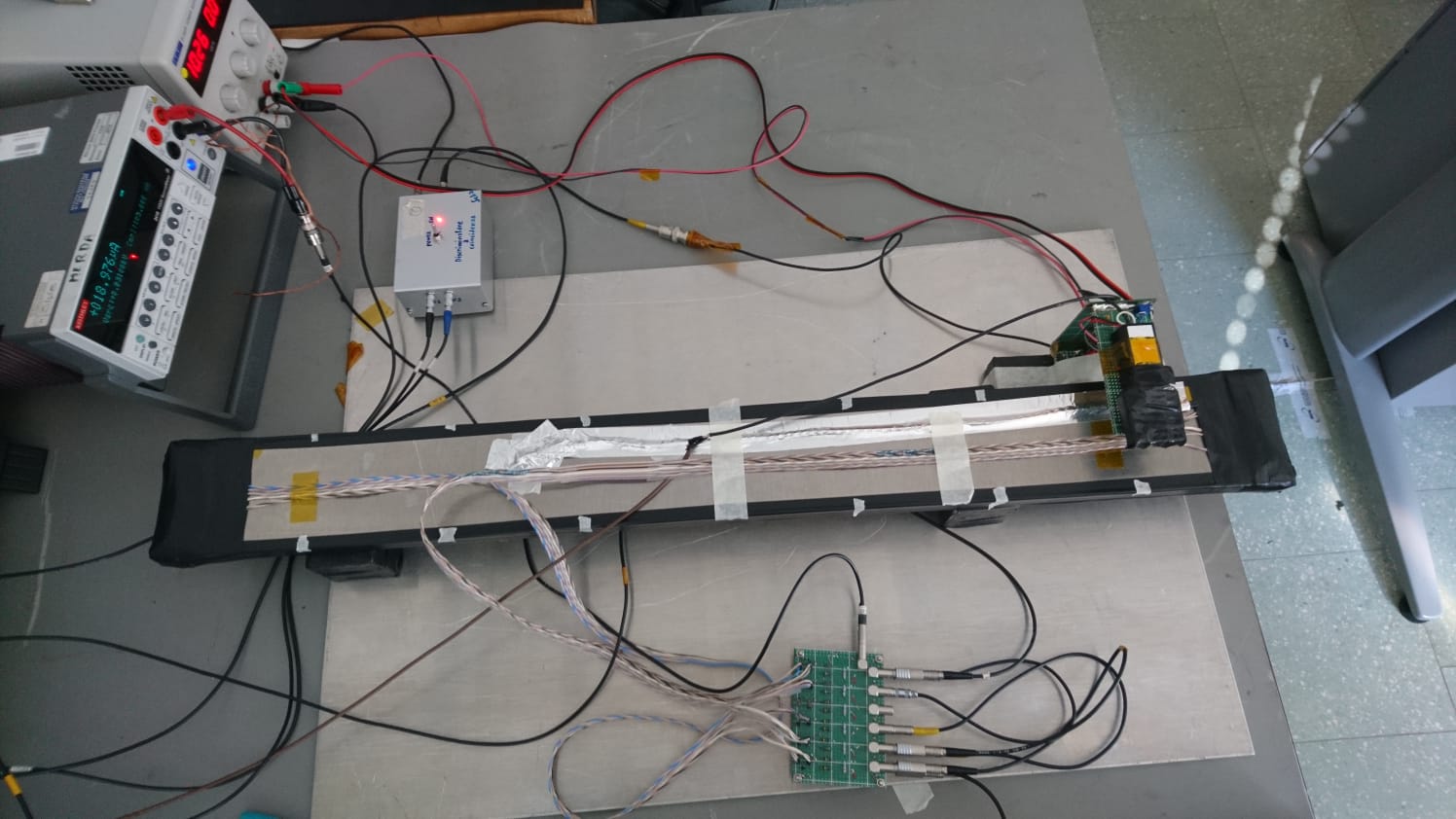}
\includegraphics[height=.21\textwidth,origin=c]{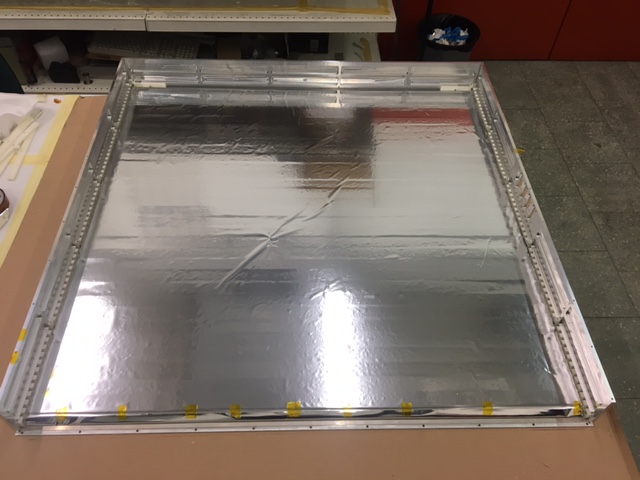}
\caption{\label{fig:prototype} Left: prototype of a partial tracking plane developed for the BLEMAB project. Right: the prototype under test: the trigger system is placed at the right end of the protection aluminum box enclosing the scintillator bars and a custom circuitry implementing the analog sum of signals from optical sensors is located in the bottom; in the top part of the figure the power supplies and the trigger coincidence unit can be seen. Right: the first full BLEMAB tracking module under construction.}
\end{figure}

A first partial prototype of the tracking plane has been produced to estimate the detection efficiency and spatial resolution that is possible to obtain in the BLEMAB detector configuration. This prototype is limited to 5 scintillator bars, assembled as shown in figure \ref{fig:prototype} (left). 

For this test we have used a LeCroy HDO6054 digital oscilloscope as DAQ system. The scintillator assembly has been enclosed in a protective aluminum box, used also as a darkening cover. Each bar is read by two 4\,mm\,$\times$\,4\,mm SiPM optical sensors whose signals are summed together. A study of the dependence of detection efficiency on the impact coordinate of muons along the bars (80\,cm length) has been carried out by using an external trigger system made of two auxiliary plastic scintillators read by means of standard PMTs. This scintillators have a small surface and are suitable to select a small region of the system under test. The complete setup can be seen in figure \ref{fig:prototype} (center). The first full BLEMAB tracking module under construction is shown at the right side in the same figure.

Data acquisition with the small prototype was repeated for seven different positions of the trigger system, with steps of 10\,cm, in order to cover the whole prototype's length. For each position a total of approximately 1000 muon events have been collected.
\begin{figure}[htbp]
\centering % \begin{center}/\end{center} takes some additional vertical space
\includegraphics[height=.3\textwidth]{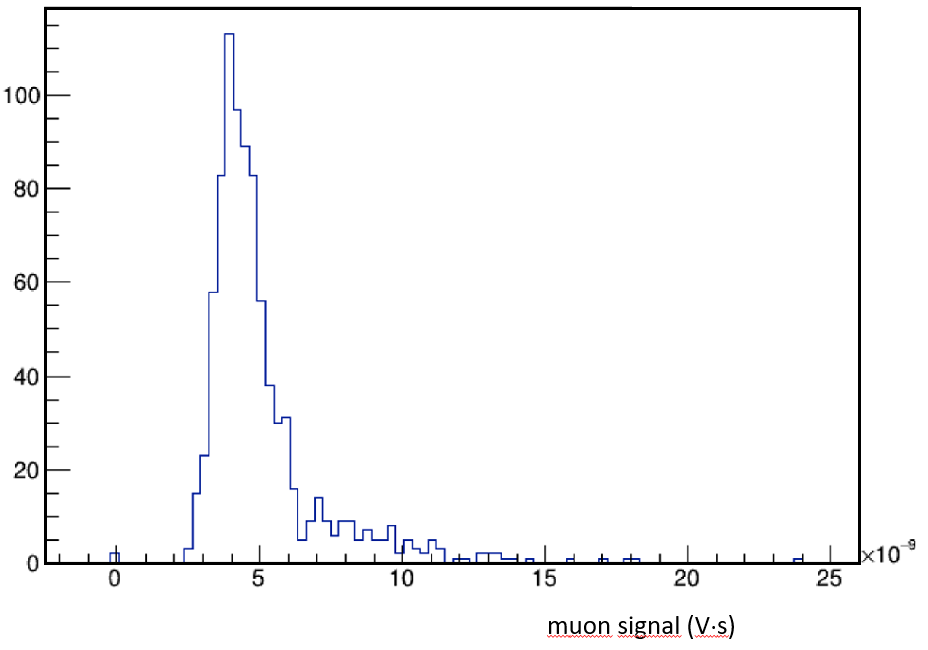}
\qquad
\includegraphics[height=.3\textwidth]{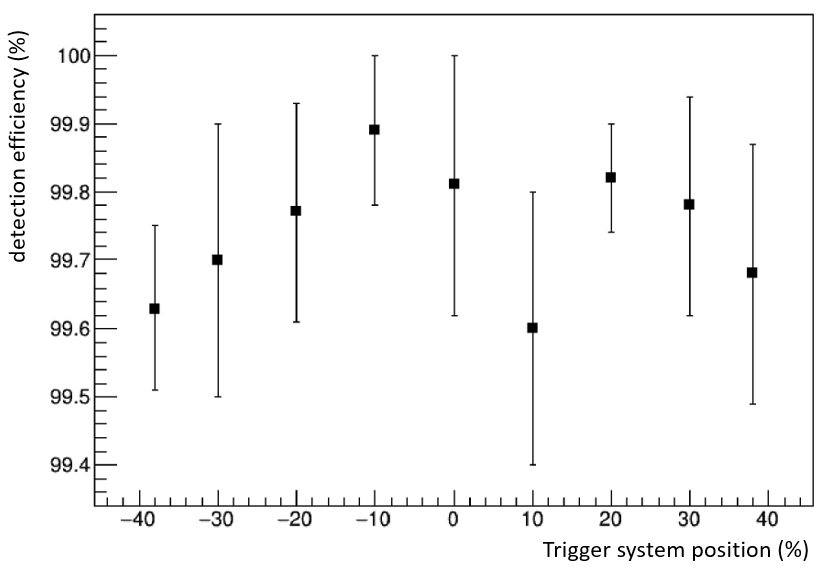}
% "\includegraphics" from the "graphicx" permits to crop (trim+clip)
% and rotate (angle) and image (and much more)
\caption{\label{fig:results} Left: distribution of signals produced by muons in the BLEMAB prototype for a fixed position of the external trigger system. Right: dependence of the detection efficiency on the muon impact point coordinate. The coordinate axis is oriented parallel to the scintillator bars.}
\end{figure}
Figure \ref{fig:results} (left) shows the histogram obtained for the total signal released by cosmic ray muons and integrated with the DAQ oscilloscope. The histogram is peaked for a value of almost 5\,V$\cdot$ns and has approximately the shape of a Landau distribution. In this measurement for 2 events of a total of 864 a null signal was registered, resulting in an efficiency of 99.8\%. Figure \ref{fig:results} (right) shows the variation of the detection efficiency for the different impact point positions. Even if some tendency could be argued, the values are all compatible with a constant value of 99.7\% within the statistical uncertainties. All measured values are beyond 99.6\%.

%********************************************************************
%********************************************************************
%**************              CONCLUSIONS               **************
%********************************************************************
%********************************************************************
\section{Final comments}
\label{sec:conclusions}
In the broad panorama of muon transmission imaging applications proposed worldwide, the European project BLEMAB represents one of the main cases of application in the industrial field. In this project the non-invasive muon transmission radiography technique will be exploited to monitor processes happening inside blast furnaces at the ArcelorMittal site in Bremen (Germany), with the main purpose of providing information about the effects of changing the blast furnace's running parameters on the distribution of materials and densities inside the blast furnace itself. In particular the geometric extension of the so called "cohesive zone", the region where the iron rich raw material begins to melt, which is characterized by a lower density with respect to the overburden material, will be studied to allow locating the interface between the two different regions. Updated measurements every few days are requested to provide useful information for the steel production process.

For this project precise software simulations and a dedicated muon tracking system suitable for the installation in the inhospitable environments of blast furnaces are under development. A small prototype of the muon tracking module has been assembled to allow a preliminary experimental validation of the project.

The final apparatus will be installed at a blast furnace in the ArcelorMittal site in Bremen in 2022. Muon imaging results will be compared with results obtained with an enhanced multi-point vertical probe and standard blast furnace models describing the variations of internal parameters along the vertical coordinate, in order to validate a new non-invasive survey methodology suitable for monitoring processes inside blast furnaces and improving their steel production efficiency.

%\appendix
%\section{Some title}
%Please always give a title also for appendices.

\acknowledgments
This project has received funding from the European Union's Research Funds for Coal and Steel 2019 Programme under grant agreement No 899263

%\paragraph{Note added.} This is also a good position for notes added
%after the paper has been written.



\begin{thebibliography}{99}
%------------------------------------------------------------------------
\bibitem{PDG}
P.A. Zyla et al. (Particle Data Group), \emph{The Review of Particle Physics}, \emph{Prog. Theor. Exp. Phys.} (2020) 083C01, direct link: https://pdg.lbl.gov/2020/reviews/rpp2020-rev-cosmic-rays.pdf
%------------------------------------------------------------------------
\bibitem{BonechiL_ADAMO_2005}
L. Bonechi et al., \emph{Development of the ADAMO detector: test with cosmic rays at different zenith angles}, \emph{29th International Cosmic Ray Conference (Pune)} {\bf 9} (2005) 283-286
%------------------------------------------------------------------------
\bibitem{SaracinoG_GALLERIABORBONICA2017}
G. Saracino et al., \emph{Imaging of underground cavities with cosmic-ray muons from observations at Mt. Echia (Naples)}, \emph{Scientific Reports} {\bf 7} (2017) 1181, DOI:10.1038/s41598-017-01277-3
%------------------------------------------------------------------------
\bibitem{BonechiL_ICNFP2017}
L. Bonechi et al., \emph{The MURAVES project and other parallel activities on muon absorption radiography}, \emph{EPJ Web Conf.} {\bf 182} (2018) 02015, DOI: 10.1051/epjconf/201818202015
%------------------------------------------------------------------------
\bibitem{BaccaniG_ICNFP2018}
G. Baccani et al., \emph{Muon Radiography of Ancient Mines: The San Silvestro Archaeo-Mining Park (Campiglia Marittima, Tuscany)}, \emph{Universe} {\bf 5(1)} (2019) 34, DOI: 10.3390/universe501003
%------------------------------------------------------------------------
\bibitem{BonechiL_IPRD2019}
L. Bonechi et al., \emph{Multidisciplinary applications of muon radiography using the MIMA detector}, \emph{JINST} {\bf 15} (2020) 05, C05030, DOI: 10.1088/1748-0221/15/05/C05030
%------------------------------------------------------------------------
\bibitem{BaccaniG_ARGINI2021}
G. Baccani et al., \emph{The reliability of muography applied in the detection of the animal burrows within River Levees validated by means of geophysical techniques}, \emph{Journal of Applied Geophysics} {\bf 191} (2021) 104376, DOI: 10.1016/j.jappgeo.2021.104376
%------------------------------------------------------------------------
\bibitem{PaganoD_ECOMUG_2021}
D. Pagano et al., \emph{EcoMug: An Efficient COsmic MUon Generator for cosmic-ray muon applications}, \emph{Nucl.Instrum.Meth.A} {\bf 1014} (2021) 165732, DOI: 10.1016/j.nima.2021.165732
%------------------------------------------------------------------------
\bibitem{GEANT4}
J.Allison et al., \emph{Recent developments in Geant4}, \emph{Nucl.Instrum.Meth.A} {\bf 835} (2016) 186-225, DOI: 10.1016/j.nima.2016.06.125
%------------------------------------------------------------------------
\bibitem{BonomiG_Review2020}
G. Bonomi et al., \emph{Applications of cosmic-ray muons}, \emph{Progress in Particle and Nuclear Physics} {\bf 112} (2020) 103768, DOI: 10.1016/j.ppnp.2020.103768
%------------------------------------------------------------------------
\bibitem{BonechiL_Review2020}
L. Bonechi et al., \emph{Atmospheric muons as an imaging tool}, \emph{Rev.Phys.} {\bf 5} (2020) 100038, DOI: 10.1016/j.revip.2020.100038
%------------------------------------------------------------------------
\bibitem{L03}
S. Zaimi et al., \emph{Blast Furnace models development and application in ArcelorMittal Group}, \emph{Rev. Met. Paris} {\bf 106 (3)} (2009) 105-111 
%------------------------------------------------------------------------
\bibitem{L04}
S. J. van der Stel et al., \emph{Stable blast furnace control by advanced measurement technique}, presentation at \emph{METEC 2015} conference (15-19 June 2015, Duesseldorf, Germany)
%------------------------------------------------------------------------
\bibitem{L05}
K. Kanbara et al., \emph{Dissection of Blast Furnaces and Their Inside State - Report on the Dissection of Blast Furnaces-1}, \emph{Tetsu-to-Hagane} {\bf 62} issue 5 (1976) 
%------------------------------------------------------------------------
%\bibitem{MUBLAST}
%E. \oAstr\"om et al., \emph{Study of the capability of muon tomography to map the material composition inside a blast furnace (Mu-blast)}, European Project RFSR-CT-2014-00027

\bibitem{MUBLAST}
E. {\AA}str{\"o}m et al., \emph{Precision measurements of Linear Scattering Density
using Muon Tomography}, \emph{JINST} {\bf 11} (2016) 07, P07010, DOI: 10.1088/1748-0221/11/07/P07010

%------------------------------------------------------------------------
\bibitem{L09}
A. Shinotake et al., \emph{Probing the inner structure of blast furnace by cosmic ray muon radiography}, \emph{Tetsu-to-Hagane} {\bf 95}  issue 10 (2009) 665-671 
%------------------------------------------------------------------------
\bibitem{DErricoME_IPRD2019}
M. D'Errico et al., \emph{Muon radiography applied to volcanoes imaging: the MURAVES experiment at Mt. Vesuvius}, \emph{JINST} {\bf 15} (2020) 03, C03014, DOI: 10.1088/1748-0221/15/03/C03014
%------------------------------------------------------------------------
\bibitem{BaccaniG_MIMA2018}
G. Baccani et al., \emph{The MIMA project. Design, construction and performances of a compact hodoscope for muon radiography applications in the context of Archaeology and geophysical prospections}, \emph{JINST} {\bf 13} (2018) 11, P11001, DOI: 10.1088/1748-0221/13/11/P11001


% Please avoid comments such as "For a review'', "For some examples",
% "and references therein" or move them in the text. In general,
% please leave only references in the bibliography and move all
% accessory text in footnotes.

% Also, please have only one work for each \bibitem.


\end{thebibliography}
\end{document}